\documentclass[12pt,twoside,a4paper]{article}
\usepackage[dvips]{epsfig}
\newcommand{\al}{\alpha}

\newcommand{\simgt}{\,\rlap{\lower 3.5 pt \hbox{$\mathchar \sim$}} \raise 1pt
 \hbox {$>$}\,}
\newcommand{\simlt}{\,\rlap{\lower 3.5 pt \hbox{$\mathchar \sim$}} \raise 1pt
 \hbox {$<$}\,}

\begin{document}
\thispagestyle{empty}
\title{\vskip-3cm{\baselineskip14pt
\centerline{\normalsize DESY 99--004 \hfill ISSN 0418--9833}
\centerline{\normalsize MPI--PhT 99--01 \hfill }
\centerline{\normalsize hep--ph/9901314 \hfill}
\centerline{\normalsize January 1999 \hfill}}
\vskip1.5cm
Forward Jet Production at small $x$ \\ in Next-to-Leading Order QCD
\author{G.~Kramer$^1$ and B.~P\"otter$^2$ \vspace{2mm} \\
$^1$ II. Institut f\"ur Theoretische Physik\thanks{Supported
by Bundesministerium f\"ur Forschung und Technologie, Bonn, Germany,
under Contract 05~7~HH~92P~(0), and by EU Fourth Framework Program
{\it Training and Mobility of Researchers} through Network {\it
Quantum Chromodynamics and Deep Structure of Elementary Particles}
under Contract FMRX--CT98--0194 (DG12 MIHT).}, Universit\"at Hamburg\\
Luruper Chaussee 149, D-22761 Hamburg, Germany \vspace{2mm} \\
$^2$ Max-Planck-Institut f\"ur Physik (Werner-Heisenberg-Institut),\\
F\"ohringer Ring 6, 80805 Munich, Germany \vspace{2mm} \\
e-mail: kramer@mail.desy.de, poetter@mppmu.mpg.de} }

\date{}
\maketitle
\begin{abstract}
\medskip
\noindent

{\parindent=0mm The} production of forward jets of transverse energy
$E_T \simeq Q$ and large momentum fraction $x_{jet} \gg x$ is calculated in
next-to-leading order including consistently direct and resolved virtual
photon contributions. The predictions are compared to recent ZEUS and H1
data. Good agreement with the data is found.
\end{abstract}

\section{Introduction}

The cross section for forward jet production in deep inelastic scattering
(DIS) has been proposed as a particularly sensitive means to investigate the
parton dynamics at small $x$ \cite{1}. Analytic calculations based on the
BFKL equation \cite{BFKL} in the leading-logarithmic approximation show
a strong rise of this cross section with decreasing $x$ \cite{2} and were
found in reasonable agreement \cite{3} with the first data from the H1
collaboration at HERA \cite{4}. More recent measurements of the forward cross
section, based on an order of magnitude increased statistics compared to 
\cite{4}, have been presented recently by the ZEUS \cite{5} and the H1 \cite{6}
collaborations confirming the earlier findings \cite{4}. Monte Carlo 
generators based on direct photon interactions (DIR) calculated from leading
order (LO) $O(\alpha_s)$ matrix elements together with leading-logarithm

parton showers disagree with the measured jet cross section \cite{5,6} by
an appreciable factor. Also next-to-leading order (NLO), i.e. $O(\alpha_s^2)$,
calculations predict too small forward cross sections at small $x$ as 
already shown by Mirkes and Zeppenfeld \cite{7} using their {\tt MEPJET} 
program \cite{14} when comparing to the data in \cite{4}. This
has been confirmed also with the new data in \cite{5,6}.

A similar deficiency between NLO calculations and measured data occurs
for the dijet rate in the region $E_T^2 > Q^2$ \cite{8}, where $E_T$
is the transverse energy of the produced jets and $Q^2$ is the usual
squared lepton momentum transfer. This kinematic range is also
relevant for the forward jet production, as will be seen later. The
region of small enough $Q^2$ is the photoproduction regime where the
virtual photon resolves into partons. Indeed, introducing a resolved 
photon contribution, the measured dijet rate and the forward jet cross section
can be described satisfactorily \cite{9,8,6} concerning the shape of the cross
section as a function of $x$ as well as the absolute normalization.
This description is based on the Monte Carlo program RAPGAP \cite{10}
which includes a resolved photon contribution in addition to the direct 
process, which both are evaluated with LO matrix elements with additional
emissions in the initial and final state generated by parton showers
together with subsequent hadronization.

The dijet rate has been calculated also in NLO including direct and
resolved photon contributions \cite{11}. In order to avoid double
counting in the full NLO calculation, the contribution from the
virtual photon splitting into $q\bar{q}$ pairs, where either the quark
or the antiquark subsequently interacts with a parton originating from
the proton, had to be subtracted \cite{12}, similar as is done in the
NLO theory for the photoproduction of jets \cite{13}. The subtracted
terms in the NLO direct contribution are part of the parton
distribution functions (PDF's) of the virtual photon and appear in the
resolved contribution in an evolved form. With this procedure the
whole cross section for two-jet production, which is a superposition
of the direct and resolved contributions minus the photon splitting
piece, becomes to a large extent independent of the factorization
scale at the photon vertex. This full NLO calculation of the dijet
rate agreed well with the H1 data over the  full $Q^2$ domain, $5 \leq
Q^2 \leq 100$~GeV$^2$, and the $x$ domain, $10^{-4} \leq x \leq 10^{-2}$,
and for jet transverse momenta $E_T^2 \geq Q^2$ \cite{6,11}.

In this work we want to present the results of a calculation of the
forward jet cross section on the basis of the NLO theory used for the dijet
rate. Although the kinematic constraints, very low $x$ and $E_T^2/Q^2$ of
order one, are rather similar, it is not obvious that the calculated cross
sections will agree with the recent ZEUS \cite{5} and H1 \cite{6} experimental 
results.

After some comparisons with the {\tt MEPJET} results \cite{7} to make sure
that our DIS jet program, called {\tt JetViP} \cite{11}, gives the same
results under identical kinematical conditions we shall give our
results with the experimental cuts of the ZEUS \cite{5} and H1
\cite{6} analysis. We close with a short summary and an outlook to
future studies.

\section{Comparisons and Results}

\subsection{Comparison with MEPJET}

Before we present our results with the ZEUS and H1 kinematical constraints
for selecting the forward jets we performed a check of our program {\tt JetViP}
with the forward jet kinematics by comparing with the NLO results of
Mirkes and Zeppenfeld \cite{7}, who have produced their results 
with the fixed order program {\tt MEPJET} \cite{14}, which only includes
direct photon contributions. We have chosen the same kinematical cuts,
which differ somewhat from the cuts used in the ZEUS \cite{5} and H1
\cite{6} analyses.

The $O(\alpha_s)$ results are obtained taking the Gl\"uck, Reya and Vogt (GRV)
LO proton PDF's \cite{15} together with the one-loop formula for $\alpha_s$.
For the $O(\alpha_s^2)$ results we employ the GRV higher order PDF's together 
with the two-loop running $\alpha_s$ formula. We take $N_f=5$ and match
the strong coupling at the charm and bottom thresholds $\mu_R=m_c,m_b$,
respectively. 

Jets are defined in the laboratory frame using the cone algorithm with
the opening angle $\sqrt{(\Delta\eta)^2+(\Delta\phi)^2}=\Delta R \leq
1$ in the so-called E-scheme. In this scheme the four-vector of the
combined jet is given as the sum of the four-vectors of the two
partons. The differences of pseudorapidities and azimuthal angles with
respect to the jet direction are $\Delta \eta$ and $\Delta \phi$. All
jets have to fulfill $|\eta|<3.5$ and $E_T,E_T^B>4$~GeV, where the
index B refers to quantities in the Breit frame. $\eta$ and $E_T$ are
measured in the HERA laboratory frame. Additional cuts are made for
events which contain a forward jet. These requirements are $1.735 < \eta
<2.90$ and $E_T>5$~GeV with 
\begin{equation}
    p_z/E_P > 0.05 \ , \qquad \qquad  0.5 < E_T^2/Q^2 < 4 \ .
\end{equation}
The $x$ variable is restricted to the small-$x$ region of $x < 0.004$. The
cuts on the electron variables are $Q^2 > 8$~GeV$^2$, $y>0.1$,
$E'>11$~GeV and $\theta'_{e} \in [160^0,173.5^0]$. The electron and
proton energies are $E_e=27.5$~GeV and $E_P=820$~GeV, respectively. The
positive $z$-direction is the direction of the incoming proton momentum.

The renormalization ($\mu_R$) and factorization scales ($\mu_F$) are taken
equal and are identified with the sum
$\mu_R=\mu_F=\frac{1}{2}\sum_{i}k_T^B(i)$ where $k_T^B(i)$ and
$p_T^B$, the parton's transverse momentum, are related in the Breit frame by
\begin{equation}
[k_T^B(i)]^2 = 2E_i^2(1-\cos\theta_{ip})=\frac{2}{1+\cos\theta_{ip}}
   [p_T^B(i)]^2  ,
\end{equation}
where $\theta_{ip}$ is the angle between the parton and the proton direction
in the Breit system. In LO one-jet production, i.e., in the naive
parton model limit, $k_T^B(i) = Q$. With these constraints we obtain
the cross sections in Tab. 1, where our {\tt JetViP} results are compared to
results from \cite{7}, referenced as {\tt MEPJET} in the table. In addition
to the forward jet cross sections we also list the full 2-jet,
inclusive 2-jet and exclusive 3-jet cross sections. Compared to these
cross sections, the  kinematic constraints defining the forward cross
section lead to considerable reductions. Furthermore we notice that
the NLO corrections to the forward jet cross sections are large in
agreement with \cite{7}. The O($\alpha_s^0$) single-jet cross section
is not considered since it vanishes if a forward jet is required, due
to the kinematical restrictions of the phase space.

\begin{table}[ttt]
\renewcommand{\arraystretch}{1.3}
\caption{Jet cross sections in the forward region compared to {\tt MEPJET}
results. \label{jv vs mep}}
\begin{center}
\begin{tabular}{lccc} \hline\hline
 Contribution & {\tt MEPJET} & {\tt JetViP} & relat.\ difference \\ \hline
 ${\cal O}(\al_s)$ 2 jet & $2120$ pb   & $2203$ pb         & $+4$\% \\
  same + forward jet & $18.9$ pb & $20.0$ pb               & $+6$ \% \\ \hline
 ${\cal O}(\al_s^2)$ 2 jet inclusive & $2400$ pb & $2371$ pb& $-1$\% \\
  same + forward jet & $83.8$ pb & $89.0$ pb               & $+6$\% \\ \hline
 ${\cal O}(\al_s^2)$ 3 jet exclusive & $210$ pb & $207$ pb & $-1$\% \\
  same + forward jet & $14.8$ pb & $14.5$ pb           & $-2$\% \\ \hline\hline
\end{tabular}
\end{center}
\end{table}

The numbers for {\tt MEPJET} are taken from ref.~\cite{7}. They differ
by a few percent from our {\tt JetViP} results. This is due to a
different implementation. In {\tt JetViP} the azimuthal ($\phi$)
dependence of the jet with respect to the electron plane is integrated
out in the hadronic center-of-mass system, whereas in {\tt MEPJET}
this $\phi$ dependence is included in this frame and  
then integrated in the HERA laboratory system. These terms which originate
from the interference of the longitudinal and transverse virtual photon
polarization ($\sim \cos \phi$) and from the transverse linear photon
polarization ($\sim \cos 2\phi$) vanish for $Q^2 \rightarrow 0$ \cite{16}.
Since in our case the virtuality $Q^2$ is not very large, the contribution
of the azimuthal dependent terms is small, which leads to small deviations
between {\tt MEPJET} and {\tt JetViP} results. In addition, when
integrating over the full phase space, it does not matter in which
system the $\phi$ integration is  performed, so that the observed
difference is essentially due to the phase space restrictions in the
forward jet selection.

Next we have evaluated the forward jet cross section including a resolved
virtual photon contribution. Since the $Q^2$ is fairly large one might
think that introducing a resolved virtual photon component is not necessary,
because it is equivalent to a contribution generated in the NLO correction
to the direct cross section. However, the NLO resolved cross section
introduces additional higher order terms that are not contained in the
NLO direct cross section, as we will see below. 
The kinematical cuts for selecting the forward 
jet are the same as for the comparison with {\tt MEPJET}. The resolved
cross section in LO and up to NLO is calculated as described in our
earlier work \cite{11} and which is incorporated in the {\tt JetViP}
program \cite{11}. The PDF's of the 
virtual photon are taken from \cite{17}, specifically we took the version
SaS1D, which was transformed to the $\overline{\mbox{MS}}$ scheme (see
\cite{11}). For the LO resolved cross section it would be more appropriate to
choose SaS1D without the transformation to the $\overline{\mbox{MS}}$ scheme.
This would increase the total LO cross section in Tab. 2 by $10\%$. 
As in \cite{11} we subtracted a term originating from the
$\gamma^{*} \to q\bar{q}$ splitting in the NLO direct photon matrix
elements in order to avoid double counting at the NLO level. The SaS
parametrizations of the virtual photon PDF vanish if the virtuality
$Q^2$ is larger than the factorization scale $\mu_F^2$. Therefore we
have chosen $\mu_R^2=\mu_F^2=Q^2+(E^B_T)^2$. This enforces the virtual
photon to be present since now always $Q^2<\mu_F^2$. The results for
the various components of the forward cross section are summarized in
Tab. 2. From these results we observe the following. First, the sum of
the LO direct and resolved contributions coincide within $10\%$ with
the NLO direct cross section. Second, adding the subtracted  NLO
direct contribution, which is the NLO direct minus the contribution
from the photon splitting term (given with the minus sign in Tab. 2)
to the NLO resolved cross section, leads to a large correction of
about $60\%$ if compared to the full NLO direct cross section. This
increase has two sources.
\begin{table}[ttt]
\renewcommand{\arraystretch}{1.3}
\caption{Resolved component in the forward region. \label{jv-res}}
\begin{center}
\begin{tabular}{lccc} \hline\hline
 Contribution & ${\cal O}(\al_s)$ 2 jet & ${\cal O}(\al_s^2)$ 2 jet
  incl. \\ \hline  
Resolved &  $61.3\pm 0.5$ pb   &  $109.6\pm 0.7$ pb \\ 
$\gamma^*\to q\bar{q}$ splitting & & $-52.9 \pm 0.3$ pb& \\
DIS Direct & $18.7\pm 0.2$ pb  & $89.6\pm 0.3$ pb  \\ \hline
Sum & $80.0\pm 0.5$ pb & $146.3\pm 0.8$ pb \\ \hline\hline
\end{tabular}
\end{center}
\end{table}
First, the LO resolved cross section is $15$\% larger than the
subtracted photon splitting term. Part of this increase is due to the
evolution of the PDF's of the photon. The other part comes from the
gluon component of the photon PDF, which amounts to 7.3 pb. Second,
the NLO corrections to the resolved cross section give a further
increase as compared to the LO result of approximately $80\%$, which
originates from the NLO corrections to the resolved matrix elements. 

We conclude, that the NLO resolved contribution supplies higher
order terms in two ways, first through the NLO corrections in the hard
scattering cross section and second in the leading logarithmic
approximation by evolving the PDF's of the virtual photon to the
chosen factorization scale. This way we sum the logarithms in
$E_T^2/Q^2$, which, however, in the considered kinematical 
region is not an important effect numerically, as we have seen in
Tab.~2. Therefore, the enhancement of the NLO direct cross section through
inclusion of resolved processes in NLO is mainly due to the
convolution of the point-like term in the photon PDF with the NLO
resolved matrix elements, which gives an approximation to the NNLO
direct cross section without resolved contributions. One of the
dominant contributions to the forward cross section is shown in Fig.~1
(left part), where the photon splitting term is convoluted with a matrix
element that provides two gluons in the final state. 
This way one gluon rung is added to the gluon ladder as compared to
the corresponding NLO direct cross section, shown on the right of
Fig.~1. The additional gluon in the NLO resolved term is in  
our approach calculated from perturbative QCD, producing an additional term
which makes a contribution to the forward jet. In the BFKL approach for
the forward jet cross section \cite{2,3} this extra gluon is part of the BFKL
evolution. In contrast, the NLO direct term in Fig. 1 (right part) contains
also additional gluons in the DGLAP evolution of the proton PDF, which, 
however, are not resolved, i.e., go to the proton remnant. Another way to
generate a larger forward jet cross section would be to go to the NNLO
corrections of the direct cross section, which has not been done yet. We
think that including the NLO resolved component produces a reasonable
approximation to this NNLO cross section. Such a correspondence is present at
one order lower. As already remarked above, the superposition of the LO
direct and resolved cross section is almost equal to the NLO direct cross
section.
\begin{figure}[ttt]
\unitlength1mm
\begin{picture}(161,53)
\put(10,-11){\psfig{file=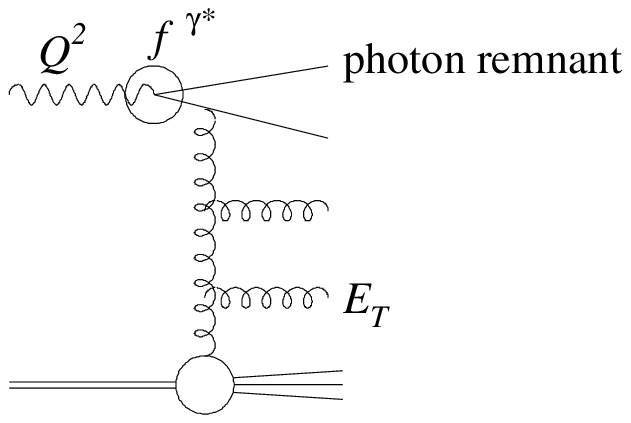,width=11cm} }
\put(80,-11){\psfig{file=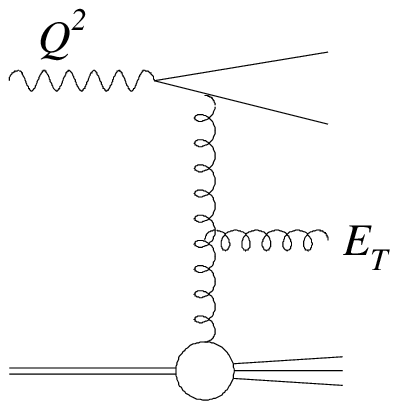,width=11cm} }
\put(0,6){\parbox[t]{16cm}{\sloppy Figure 1: Diagrams contributing
        to resolved (left) and direct (right) processes in the forward
        region.}}
\end{picture}
\end{figure}

\subsection{Comparison with ZEUS and H1 Data}

For the comparison with the 1995 ZEUS \cite{5} and the 1994 H1 \cite{6}
forward jet cross section data we calculated the NLO cross section
with slightly different input and in particular with the exact kinematical
constraints for the forward jet selection as used in the two experiments.

As proton PDF's we apply now the CTEQ4M parametrization \cite{18} with 
the two-loop $\alpha_s$. We take $N_f=5$ as before and match the value of
$\alpha_s$ at the thresholds $\mu_R=m_c,m_b$ with a $\Lambda_{\overline{MS}}$
as used in CTEQ4M. Jets are defined with the cone algorithm in the HERA frame
as described above, except that the axis of the jet is calculated now as the
transverse energy weighted mean of $\eta$ and $\phi$ of the two partons or
jets belonging to the combined jet. This kind of jet definition was also 
applied in the experimental jet analysis. As scales we choose 
$\mu^2=\mu_R^2=\mu_F^2=M^2+Q^2$ with a fixed $M^2=50$~GeV$^2$ related to the
mean $E_T^2$ of the forward jet. We take this fixed value of $M$ instead of 
$E_T$ for technical reasons, since the calculations in {\tt JetViP}
start from the hadronic c.m.s.. The choice $\mu_F^2 > Q^2$ is
mandatory if we want to include a resolved contribution. Another choice of
scale would be $\mu_F=E_T$ or $\mu_F=M$. In the case $\mu_F^2/Q^2 > 1$ only
for $E_T^2/Q^2 > 1$, which covers only part of the ZEUS kinematical range.
To have a resolved cross section in all $E_T^2/Q^2$ bins we consider the
choice $\mu_F^2=M^2+Q^2$ more appropriate. 

In the two experiments the forward jet selection criteria are different. In 
the ZEUS experiment the kinematical constraints are: $E'_e>10$~GeV, $y>0.1$,
$\eta > 2.6$ and $E_T>5$~GeV with the forward jet constraints
$x_{jet}=E_{jet}/E_P > 0.036$, $0.5< E_T^2/Q^2< 2$, $p^B_{z,jet}>0$
and $4.5 \times 10^{-4} <x< 4.5 \times 10^{-2}$.
Our results for the forward jet cross section under these ZEUS kinematical
conditions are shown in Fig. 2 a,b. In Fig. 2 a we plotted the full 
O($\alpha_s^2$) inclusive two-jet cross section (DIR) as a function of $x$
for three different scales $\mu^2=3M^2+Q^2, M^2+Q^2$ and $M^2/3+Q^2$
and compared them with the measured points from ZEUS \cite{5}. As to
be expected the calculated NLO direct cross section is by a factor 2
to 4 too small compared to the data. The variation inside the assumed range
of scales is small, so that also with a reasonable change of scales we
can not get agreement with the data. In Fig. 2 b we show the
corresponding forward jet cross sections with the NLO resolved
contribution included, as described in the previous subsection, again
for the  three different scales $\mu $ as in Fig. 2 a. Now we find
good agreement with  the ZEUS data. The scale variation of the
calculated cross section is larger than in Fig. 2 a. In particular,
the largest scale gives now the largest cross section opposite to what
is observed in Fig. 2 a. This different scale 
variation comes primarily from the scale dependence of the virtual
photon PDF. This factorization scale variation is supposed to be
compensated between the LO resolved and the NLO direct contribution
but not for the NLO resolved contribution. This could only occur if we
could include the NNLO direct contribution which, however, is not
available. Since the NLO resolved contribution for the forward jet is
rather large, as discussed above, this scale dependence can not be
avoided. On the other hand, the scale dependence is not so large that
we must fear our results not to be trustworthy. In Fig. 2 b the cross
section is labeled DIR$_S$+RES, where DIR$_S$ stands for NLO direct minus
the photon-quark-antiquark splitting term. We have calculated the forward
jet cross section also with the scale $\mu_F^2=M^2$. For this choice 
we obtain a somewhat smaller cross section which is approximately equal to the 
cross section with the scale $\mu_F^2 = M^2/3+Q^2$.
\begin{figure}[ttt]
  \unitlength1mm
  \begin{picture}(122,70)
    \put(-6,-60){\epsfig{file=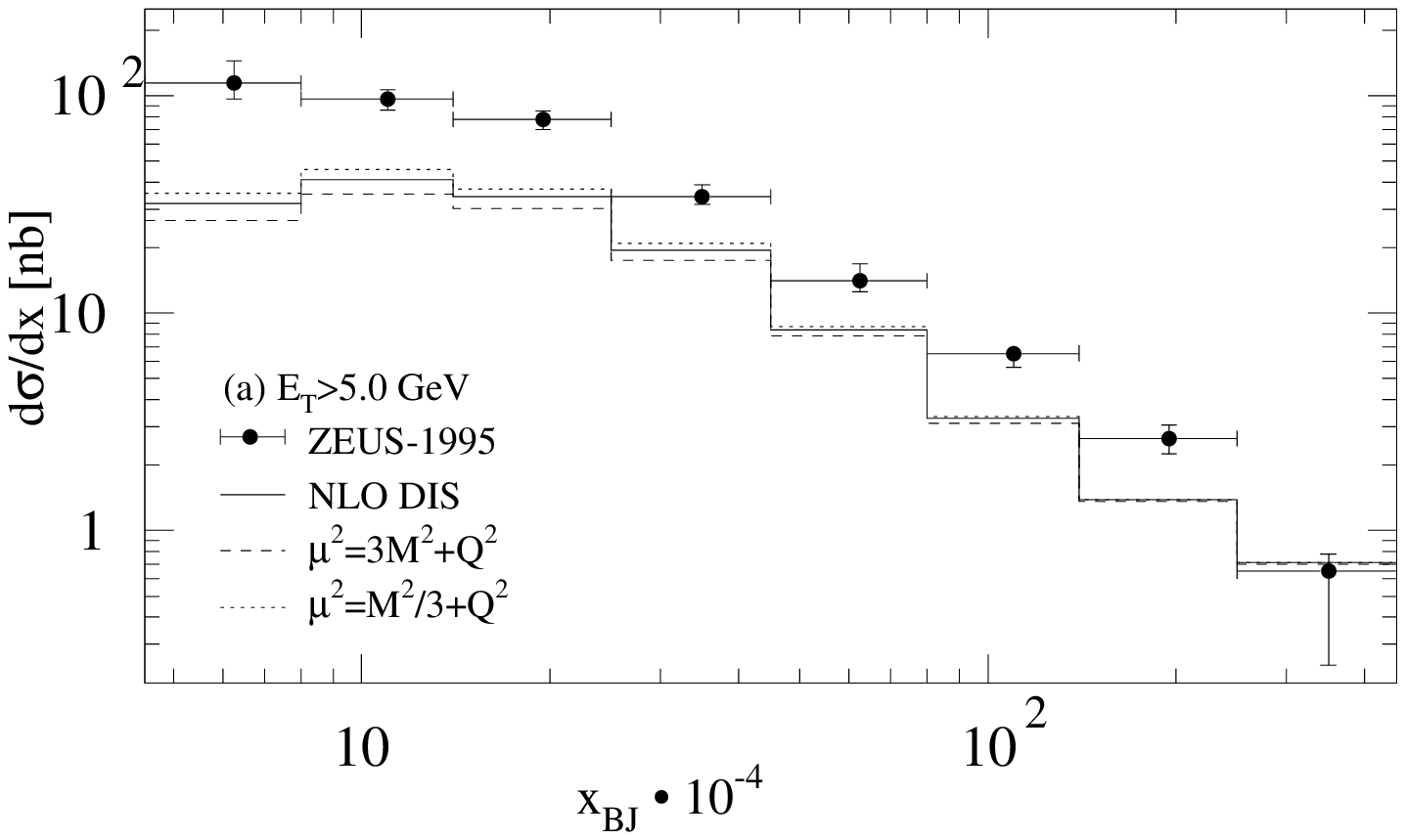,width=9.5cm,height=14cm}}
    \put(80,-60){\epsfig{file=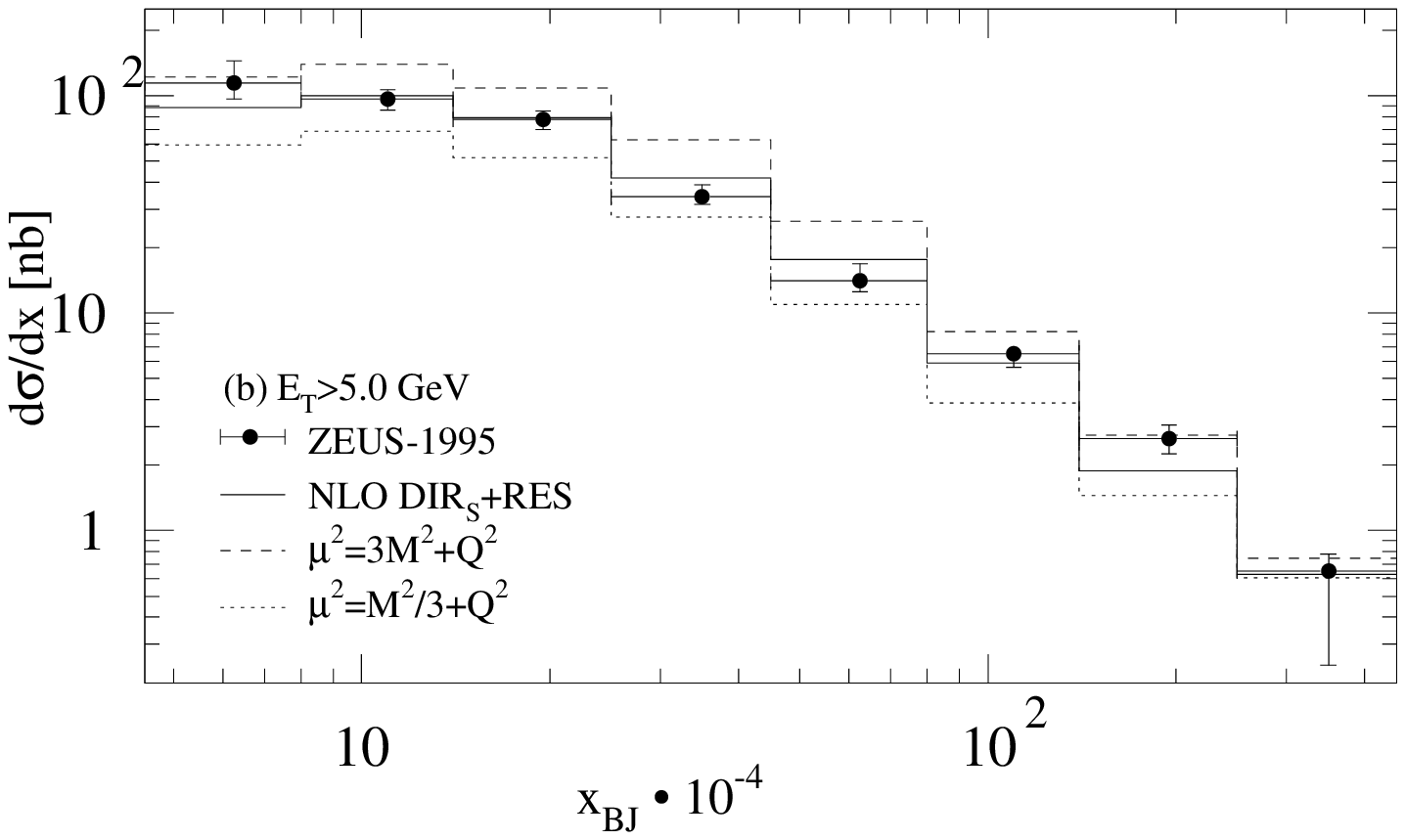,width=9.5cm,height=14cm}}
    \put(0,0){\parbox[t]{16cm}{\sloppy Figure 2: Dijet cross section
          in the forward region compared to ZEUS data. (a) NLO DIS,
          $E_T>5$ GeV (b) NLO DIR$_S$+RES, $E_T>5$ GeV.}}
  \end{picture}
\end{figure}

\begin{figure}[hhh]
  \unitlength1mm
  \begin{picture}(122,140)
    \put(-6,10){\epsfig{file=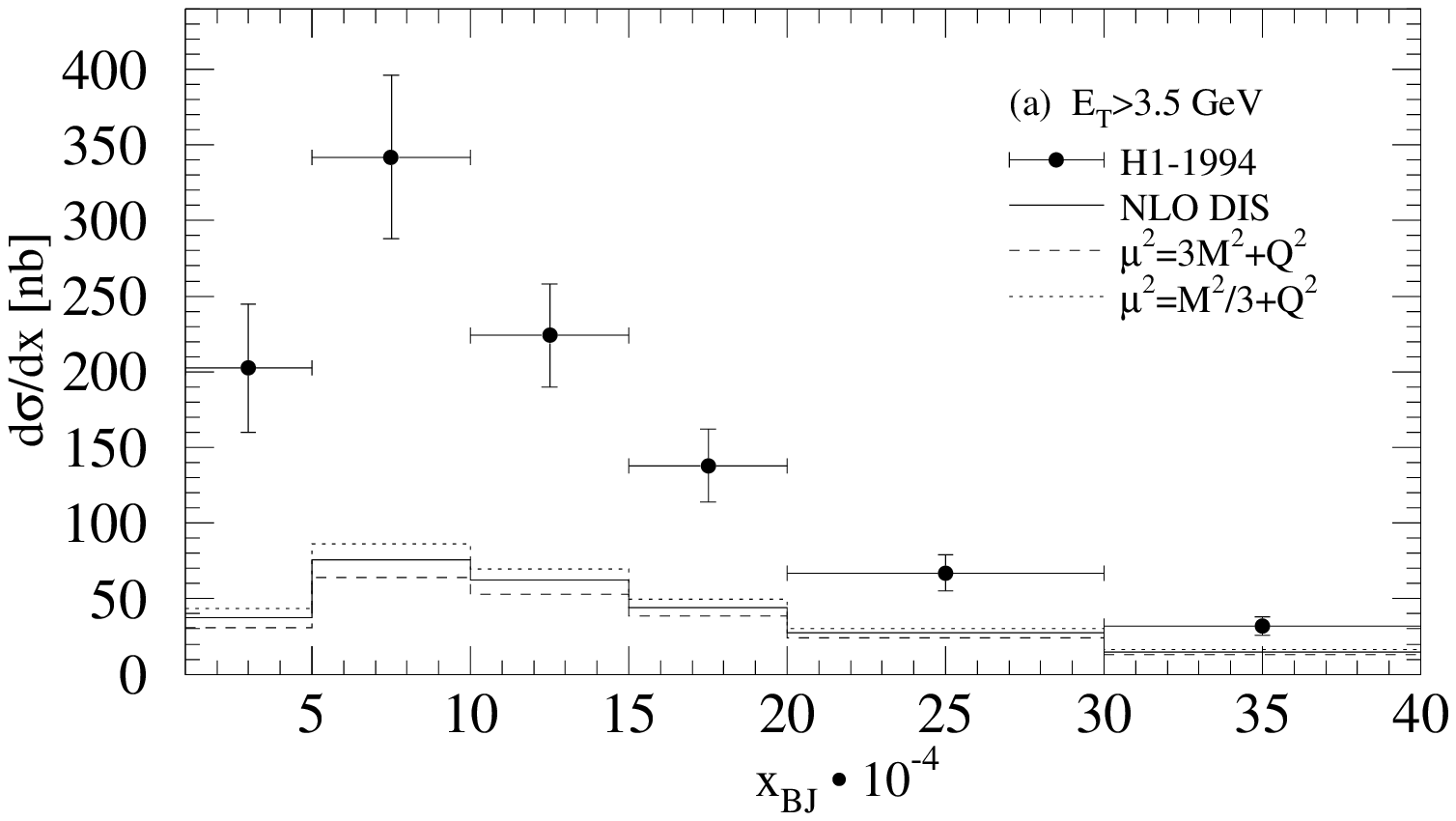,width=9.5cm,height=14cm}}
    \put(80,10){\epsfig{file=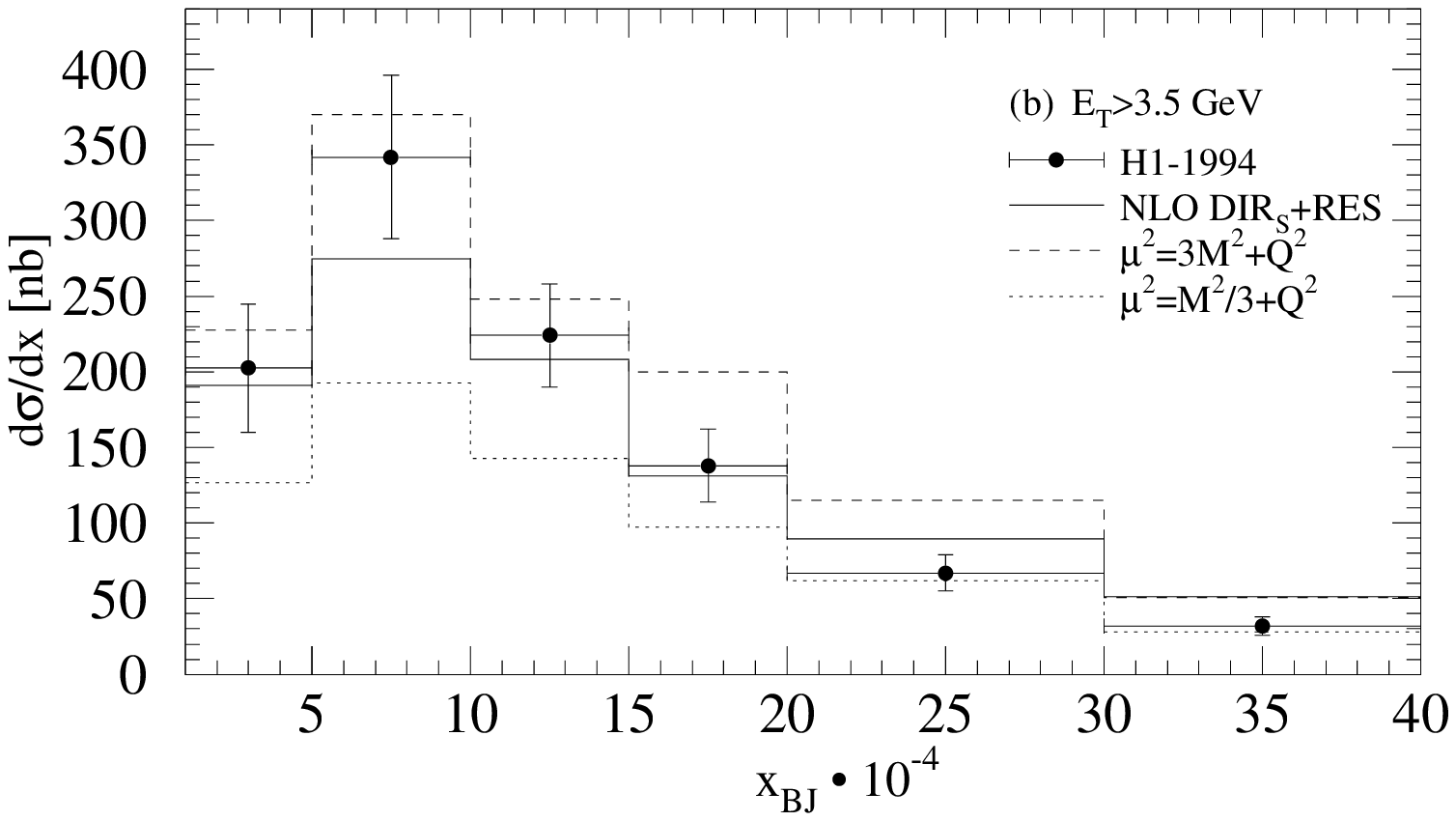,width=9.5cm,height=14cm}}
    \put(-6,-50){\epsfig{file=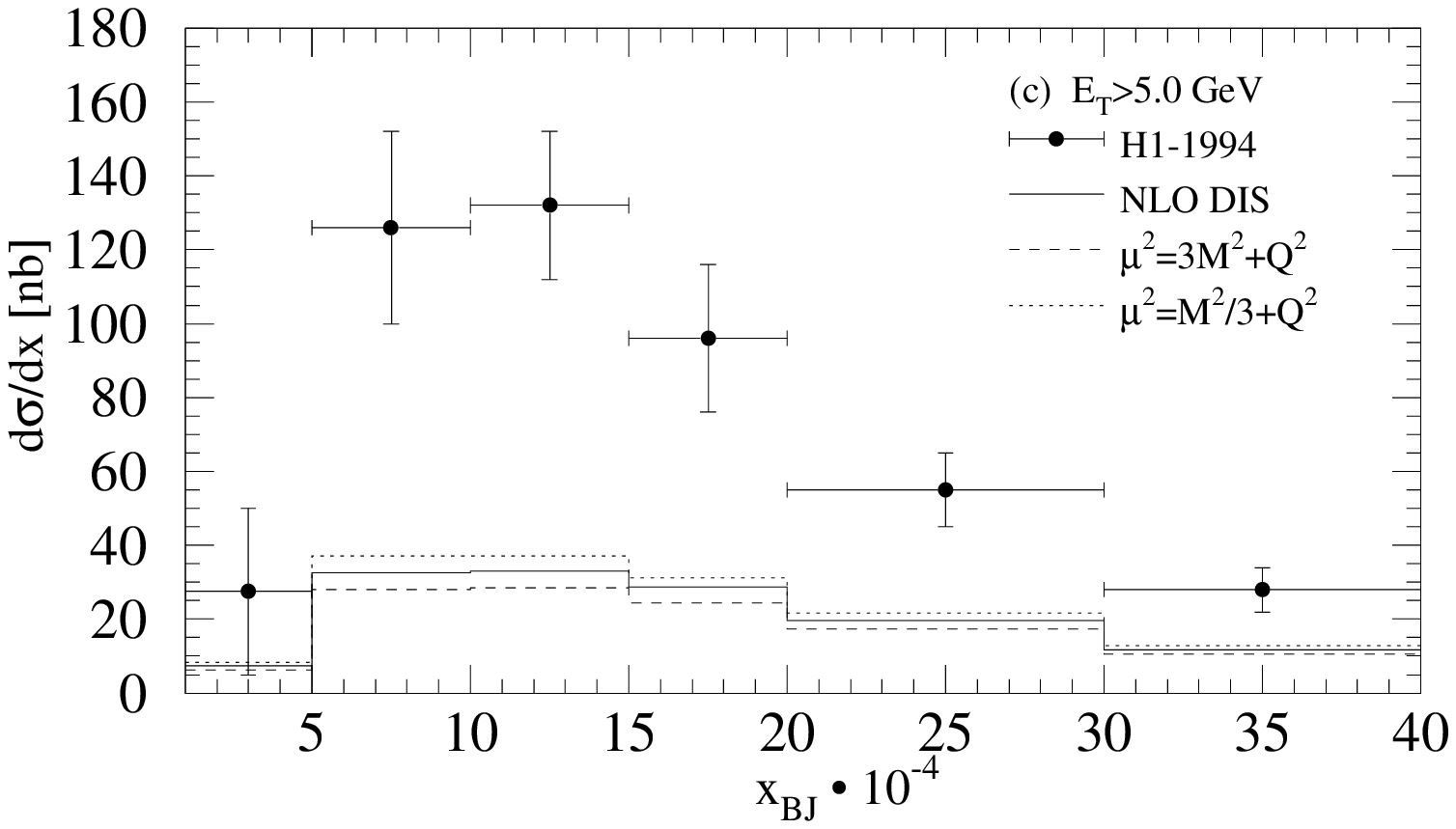,width=9.5cm,height=14cm}}
    \put(80,-50){\epsfig{file=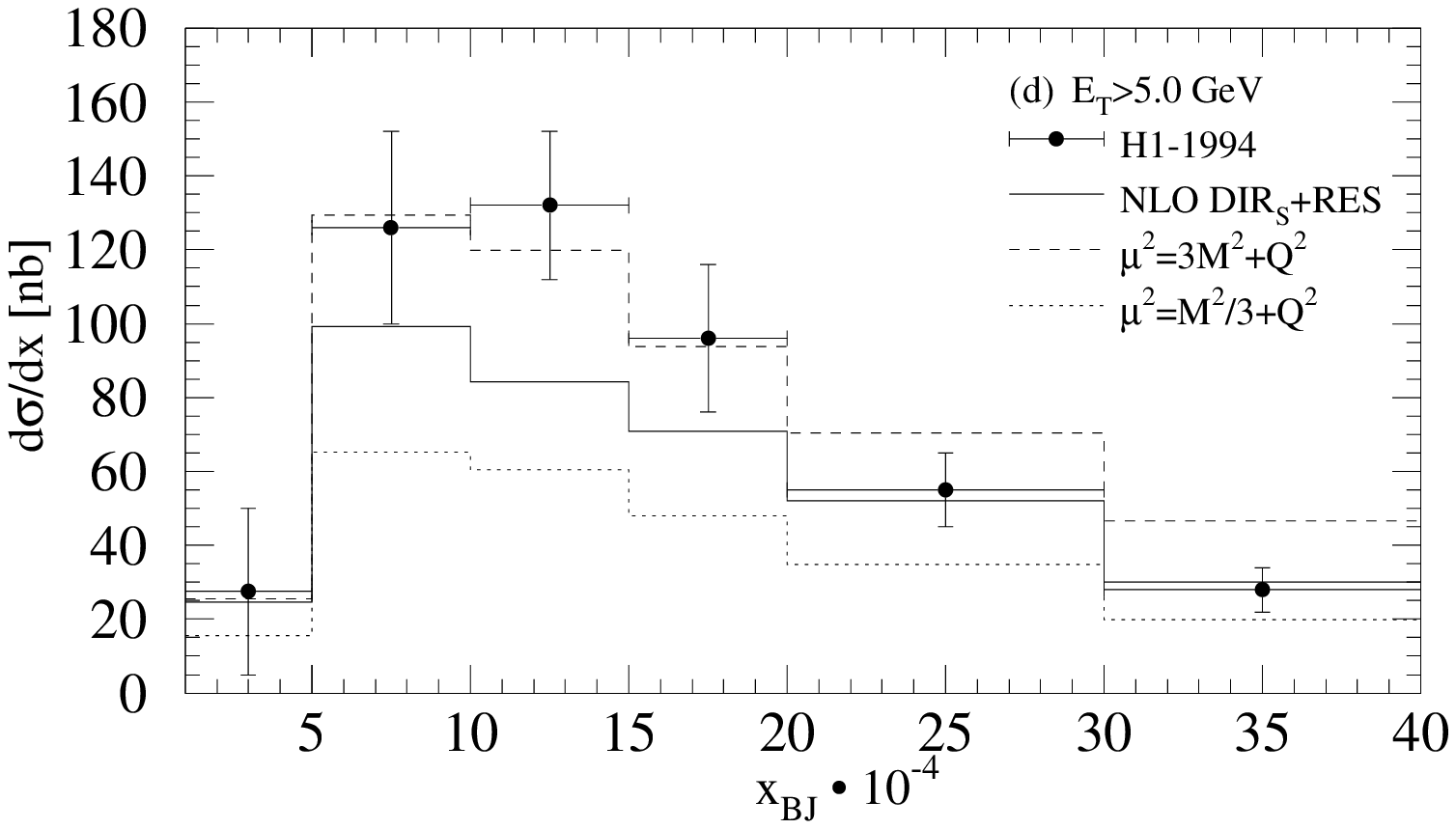,width=9.5cm,height=14cm}}
    \put(0,10){\parbox[t]{16cm}{\sloppy Figure 3: Dijet cross section
          in the forward region compared to H1 data. 
         (a) NLO DIS, $E_T>3.5$ GeV (b) NLO
          DIR$_S$+RES, $E_T>3.5$ GeV; (c) NLO DIS, $E_T>5.0$ GeV (d) NLO
          DIR$_S$+RES, $E_T>5.0$ GeV.}}
  \end{picture}
\end{figure}
In the H1 experiment \cite{6} the forward jets are selected with the 
kinematical cuts $E'_e>11$~GeV, $y>0.1$, $160^0< \theta '_e < 173^0$, 
$1.735 < \eta < 2.794$ (this corresponds to $7^0 < \theta_{jet} < 20^0$),
$E_T > 3.5~ (5.0)$~GeV, $x_{jet}>0.035$ and $0.5< E_T^2/Q^2 <2$. This
corresponds approximately to the $Q^2$ range $5<Q^2<100$~GeV$^2$. The
forward jet cross section is measured for various $x$ bins ranging
from $1.0 \times 10^{-4}$ to $4.0 \times 10^{-3}$. Otherwise the
calculated forward cross section is obtained under the same
assumptions as for the ZEUS selection cuts. In Fig. 3 a,b,c,d we show
the results compared to the H1 data obtained with two $E_T$ cuts in
the HERA system, $E_T>3.5$~GeV (Fig. 3 a,b) and $E_T>5.0$~GeV (Fig. 3
c,d). In the plots on the left (Fig. 3 a,c) the data are compared with
the pure NLO direct prediction, which turns out to be too small by a
similar factor as observed in the comparison with the ZEUS data. In
Fig. 3 b,c the forward jet cross section is plotted with the NLO
resolved contribution included in the way described above. For both
$E_T$ cuts, $E_T>3.5$~GeV (in Fig. 3b) and $E_T>5.0$~GeV (Fig. 3d), we
find good agreement with the 1994 H1 data inside the scale
variation window $M^2/3+Q^2 < \mu^2 < 3M^2+Q^2$. Compared to the ZEUS
data the H1 measurements extend down to smaller $x$. In the lowest
$x$ bin, $1.0 \times 10^{-4} < x < 5.0 \times 10^{-4}$ the forward jet
cross  section has a dip, which is also reproduced in the theoretical
calculation and which is due to the kinematical constraints for
selecting forward jets.

We conclude that the NLO theory with a resolved virtual photon contribution
gives a good description of both, the ZEUS and the H1 forward jet
data. It is important, that both components, the direct and the
resolved one, are calculated up to NLO. A LO calculation of both
components would fall short of the experimental data, as it is clear
from the results presented in Tab. 2, where we compared LO and NLO results.

We remark that the forward jet cross sections as measured by ZEUS and H1 are
obtained at the hadron level, i.e., the jets are constructed from measured
hadron momenta using the same cone algorithm. In our NLO calculation the jets
are combined from partons with the same jet algorithm. The size of the 
corrections from hadron to parton level has been studied by the ZEUS 
collaboration \cite{5} using several Monte Carlo simulation programs with the
result that for the models which account very well for the ZEUS forward
jet cross sections the correction factors are close to unity for all $x$
values considered in the analysis. In the H1 work similar results are 
reported \cite{6}.

In \cite{6} the H1 forward jet data are also successfully described with the
RAPGAP model \cite{10} which includes a direct and a resolved component,
both in LO, and with a similar scale $\mu $ as we have used. In addition
this model has leading logarithm parton showers in the initial and final state
built in. We think that these parton shower contributions produce the higher
order effects which we found necessary to account for the correct
normalization and the $x$ dependence of the forward jet cross section data.

\section{Concluding Remarks} 

We conclude that the measurements of the forward jet cross section
presented recently by the ZEUS and H1 collaborations can be described
very well by the NLO theory with direct and resolved virtual photon
contributions added in a consistent way. The theory shows good
agreement with the data with respect to the normalization and also the
functional dependence with decreasing $x$.  Whereas this variation of
the cross section with $x$ is also compatible with the NLO predictions
based on direct photons, for the correct normalization the resolved
component up to NLO is needed. To avoid double counting the 
$\gamma^*\to q\bar{q}$ splitting term is removed from the NLO direct
contribution.

In contrast, LO BFKL predictions \cite{4} yield much larger forward cross  
sections than the data \cite{5,6}. These calculations suffer, however,
from several deficiencies. They are asymptotic and do not contain the
correct kinematic constraints of the produced jets
\cite{19}. Furthermore they do not allow the implementation of a jet
algorithm as used in the experimental analysis. Also NLO $\ln(1/x)$
terms in the BFKL kernel \cite{20} predict large negative corrections
which are expected to reduce the forward cross section as well. When
all these points are taken into account the BFKL approach may give an
equally well description of the forward jet data. Even if this is the
case, it is clear from this work that the BFKL theory is not the only
theoretical approach that describes the forward jet cross sections.

\subsection*{Acknowledgements}

We are grateful to D.~Graudenz, G.~Grindhammer and H.~Jung for
interesting discussions and to D. Zeppenfeld for correspondence about
his work with E.~Mirkes.


\end{document}